# Towards Accurate and Scalable High-throughput MOF Adsorption Screening: Merging Classical Force Fields and Universal Machine Learned Interatomic Potentials

Satyanarayana Bonakala[1], Mohammad Wahiduzzaman[1], Taku Watanabe[2], Karim Hamzaoui[2], and Guillaume Maurin[1,3]*

[1]*ICGM, Univ. Montpellier, CNRS, ENSCM, F-34095 Montpellier, France*
[2]*Matlantis Corporation, Inc. Otemachi, Chiyoda-ku, Tokyo 100-0004, Japan*

[3]*Institut Universitaire de France, France*

## Abstract

High-throughput computational screening (HTCS) of gas adsorption in metal–organic frameworks (MOFs) typically relies on classical generic force fields such as Universal Force Field (UFF) which are computationally efficient but often fail to capture complex host–guest interactions. Universal machine learned interatomic potentials (u-MLIPs) offer near–quantum accuracy at far lower cost than density-functional theory (DFT), yet their large-scale application in adsorption screening remains limited. Here, we present a hybrid screening strategy that merges Widom insertion Monte Carlo simulations performed with both UFF and PreFerred Potential (PFP) u-MLIP to evaluate the adsorption performance of a large MOF structure database, using ethylene capture under humid conditions as a benchmark. From a curated set of MOFs, 88 promising candidates initially identified using UFF-based HTCS were re-evaluated with the PFP u-MLIP, benchmarked against DFT calculations, to refine the adsorption predictions and to assess the role of MOF framework flexibility. We demonstrated that PFP u-MLIP is essential to accurately assess the sorption performance of MOFs involving strong hydrogen bonding or confinement pockets within narrow pores—effects hardly accurately captured using UFF. Notably, accounting for the framework flexibility though full unit cell relaxation led to deviations in ethylene affinity of up to 20 kJ mol$^{-1}$, highlighting the critical influence of guest-induced structural changes. This overall HTCS workflow identified seven MOFs with optimal pore sizes, high ethylene affinity, and high $C_2H_4/H_2O$ selectivity, offering moisture-tolerant performance for applications ranging from food packaging to trace ethylene removal. Our findings emphasize the importance of accurately capturing host-guest energetics and framework flexibility, and demonstrate the practicality of incorporating u-MLIPs into scalable high-throughput workflows for the identification of the best MOF sorbents for a given adsorption-related application.

## I. Introduction

Metal–Organic Frameworks (MOFs), a class of hybrid porous materials with remarkable chemical and structural diversity, hold immense promise across a wide range of applications. Yet, their full potential, particularly for sorption-based processes, remains largely untapped. One key reason is that only a small fraction of the 128,000+ MOF structures reported in the Cambridge Structural Database (CSD)[1] have been explored for specific applications. This limited exploration is primarily due to the impracticality of synthesizing and experimentally testing thousands of candidates, given the substantial investment of time, resources and specialized infrastructure required. To overcome these limitations, computational approaches have emerged as indispensable tools. In particular, high-throughput computational screening (HTCS) has revolutionized the evaluation of MOFs and related porous materials properties across large structural databases. This strategy enables the rapid identification of promising sorbent candidates for a wide array of adsorption/separation applications, including carbon capture, $H_2$ storage, natural gas purification, among others.[2–6] Traditionally, these computational screenings have relied on classical interatomic potentials, commonly referred to as classical universal interatomic potentials (u-IPs) to describe the MOF/guest interactions. The most widely used classical u-IPs, e.g., UFF[7] employ transferable Lennard-Jones parameters to account for van der Waals interactions, complemented by electrostatic terms to describe interatomic interactions within the system.

While this modeling approach enables an effective qualitative assessment of thousands of MOF sorbents, it inherently lacks the precision required to capture complex host-guest interactions, especially between guest molecules and functionalized linkers as well as open metal sites (OMS) and polar groups within the inorganic nodes, where polarization effect may play a dominant role.[9–11] For example, UFF significantly underestimates $CO_2$ interaction energies with the OMSs of Mg-MOF-74 by 60-65% compared to density-functional theory (DFT) calculations,[12] which can potentially lead to the mis-ranking of this sub-class of MOFs. To overcome these shortcomings, case-specific FF parametrization has been developed using exhaustive DFT-derived interaction energies for individual MOF-guest pairs, e.g., $CO_2$@Mg-MOF-74,[13] $C_2H_2$@Cu-BTC,[14] $N_2$@MIL-100(Cr)[15] and $SO_2$@$Mg_2$(dobpdc).[16] More recently, Machine-learned potentials (MLPs), trained on high-quality quantum mechanical datasets, have proven effective in capturing intricate host/guest interactions across a range of MOFs, achieving with near quantum mechanical-accuracy at significantly lower computational cost.[17–26] For instance, this strategy has been successfully applied to accurately predict the $H_2$ adsorption isotherm in the low-pressure domain of Al-*soc*-MOF-**1d** containing OMSs.[26] Despite their improved accuracy in targeted studies for a few MOFs, both case-specific FFs and trained MLPs remain limited in transferability and are computationally impractical for large-scale screenings. In this context, the on-going development of universal machine learned interatomic potentials (u-MLIPs) offers a potentially transformative advance. Foundation models such as the Multi-Atomic Cluster Expansion (MACE),[27,28] Materials Graph Network with three-body interactions (M3GNet)[29,30] and Crystal Hamiltonian Graph Neural Network (CHGNet)[31] architectures have demonstrated near-DFT accuracy for predicting MOF framework properties offering a promising pathway towards MLP-based generalizable MOF screening platforms. Notably, Meta's Equiformer V2 trained on the Open DAC datasets, ODAC23 and ODAC25[32] have shown superior performance over the UFF in predicting $CO_2$ and $H_2O$ adsorption energies for a few MOFs, especially those featuring OMSs.[24,33] A few recent benchmark studies extended this comparison across a small sub-set of MOFs evaluating a range of u-MLIPs including EqV2-ODAC,[33] MACE,[34] M3GNet,[29] CHGNet,[31] and the Equivariant Smooth Energy Network (eSEN)[35] against DFT and UFF calculations for computing their $CO_2$ and $H_2O$ adsorption energies.[24,36,37]

However, despite these promising developments, u-MLIPs have not yet been systematically applied to predict adsorption properties across a large and diverse set of MOFs. Their integration into HTCS pipelines is still not feasible due to computational costs, which, while significantly lower compared to DFT, are still substantially higher than those associated with classical FFs. Herein, we advance a hybrid HTCS workflow for MOFs that bridges classical u-IPs and u-MLIPs, using ethylene ($C_2H_4$) capture under humid conditions as a test case. This adsorption scenario poses a fundamental challenge: weakly physisorbed ethylene must compete with strongly interacting polar water molecules, requiring accurate modelling of the MOF/guests interactions. Ethylene was selected due to its high-impact relevance in food preservation technologies, particularly in regulating the ripening of fruits and vegetables.[38] Its uncontrolled accumulation in postharvest storage and food packaging, environments typically characterized by high humidity,[39] can accelerate spoilage, contributing significantly to food loss.[40] A few technologies have been proposed to mitigate ethylene accumulation, including chemical inhibitors such as 1-methylcyclopropene (1-MCP),[41–43] oxidative scavenging agents like potassium permanganate ($KMnO_4$)[44–46] and $TiO_2$-based photocatalysts.[47] A truly viable sorbent-based solution has yet to be realized, as conventional porous materials, such as zeolites and activated carbons, exhibit inherently low $C_2H_4$ affinity and/or inadequate $C_2H_4/H_2O$ selectivity.[48–50] Recently, a few biodegradable and non-toxic MOFs have been explored as alternative sorbents for food-related applications.[51–54] For example, cyclodextrin-based MOFs (CD-MOF-Na/K),[54] synthesized from bio-derived α-cyclodextrins and alkali metal ions, have been proposed as effective $C_2H_4$ sorbents for delaying the ripening of fruits under ambient conditions. These preliminary findings highlight the potential of MOFs in selective $C_2H_4$ capture and pave the way for the identification of other existing MOFs with optimized $C_2H_4$ sorption properties.

Our proposed computational workflow begins with a HTCS of a curated database of sustainable MOFs selected from the 128,799 experimentally synthesized MOFs in the latest release of the Cambridge Structure Database.[1,55] Following a rigorous MOF structure curation and geometry optimization pipeline, key adsorption descriptors, Henry's law constants ($K_H$) and enthalpy of adsorption at infinite dilution ($\Delta H_{0,ads}$) are computed via Widom insertion method at room temperature for both $C_2H_4$ and $H_2O$ as well as the resulting ideal selectivity ($S_{C_2H_4/H_2O} = K_{H,C_2H_4}/K_{H,H_2O}$) using the UFF. This first conventional stage enables the identification of hydrophobic MOF candidates capable of selectively capturing ethylene over water. The Preferred Potential (PFP) u-MLIP, an equivariant graph neural network AI-model trained on a comprehensive dataset of 42 million DFT-calculated structures (including molecules, crystals, slabs, clusters and disordered systems),[56] is subsequently employed to refine the performance predictions of the 88 experimentally viable ethylene selective MOFs identified in the initial screening. Following its rigorous validation against DFT-calculated adsorption energies for water and ethylene across these 88 MOFs, the PFP u-MLIP is systematically applied to compute $\Delta H_{0,ads}$ and $K_H$ for both guest molecules and the corresponding $S(C_2H_4/H_2O)$. Notably, under the commonly adopted rigid-MOF framework approximation in HTCS workflows, the UFF approach demonstrates reliable performance in capturing the hydrophobic behaviour of most MOFs examined in this study, and in yielding ethylene adsorption enthalpies that are comparable to those predicted by the PFP u-MLIP model with a mean absolute deviation (MAD) of 5 kJ mol$^{-1}$. The use of PFP u-MLIP becomes more critical for a small number of outlier MOFs featuring small pore sizes or polar functional groups, where UFF proves less effective at accurately capturing $C_2H_4$ and $H_2O$ adsorption energetics, particularly under high confinement or when hydrogen-bonding interactions play a significant role. While UFF performs adequately for most cases, these outliers highlight the limitations of classical generic FFs and the advantages of higher-fidelity models like PFP u-MLIP. Notably, PFP u-MLIP results show that 99% of the MOFs exhibit ideal selectivity, $S(C_2H_4/H_2O) > 1$, confirming their strong ethylene preference and suitability for separation under humid conditions.

Although guest-triggered flexibility in MOFs is well documented and has been shown to significantly affect host-guest interactions in specific cases,[24] incorporating such structural responsiveness into HTCS workflows remains computationally prohibitive. Assessing whether the rigid-framework approximation introduces systematic deviations or remains within acceptable error margins is critical to ensuring the reliability of HTCS-derived adsorption properties. To this end, the PFP u-MLIP model was also employed to refine the predictions of $\Delta H_{0,ads}(C_2H_4)$ for the 88 top MOF candidates, explicitly accounting for adsorption-induced framework flexibility, including atomic displacements alone or in combination with unit cell relaxations. It was found that for most MOFs exhibiting guest-induced unit cell volume changes below 10% led to only moderate deviations in $\Delta H_{0,ads}(C_2H_4)$ relative to the rigid-framework approximation (mean absolute deviation (MAD) of 1.5 kJ mol$^{-1}$). This refined step enables to exclude a few otherwise attractive MOFs that undergo significantly larger guest-induced unit cell expansion, accompanied by a notable decrease in $C_2H_4$ affinity.

Beyond introducing a robust data-driven strategy for identifying MOFs suitable for ethylene removal under humid food storage conditions as a test case, this work proposes a generalized hybrid HTCS workflow that balances computational efficiency with predictive accuracy. We show that classical generic FF, UFF combined with rigid framework approximations provide a reliable and scalable first-pass filter to identify top-performing MOF candidates for sorption-based applications. For the top-ranked MOFs, we then leverage PFP u-MLIP to refine UFF-derived performance predictions, capturing host/guest interactions and structural flexibility more accurately. Our findings provide valuable insights into the transformative potential of u-MLIPs in HTCS, offering a powerful strategy to refine classical u-IPs predictions and to enable more accurate and resource-efficient screening pipelines for MOF-based sorption properties.

## II. Computational Methodology

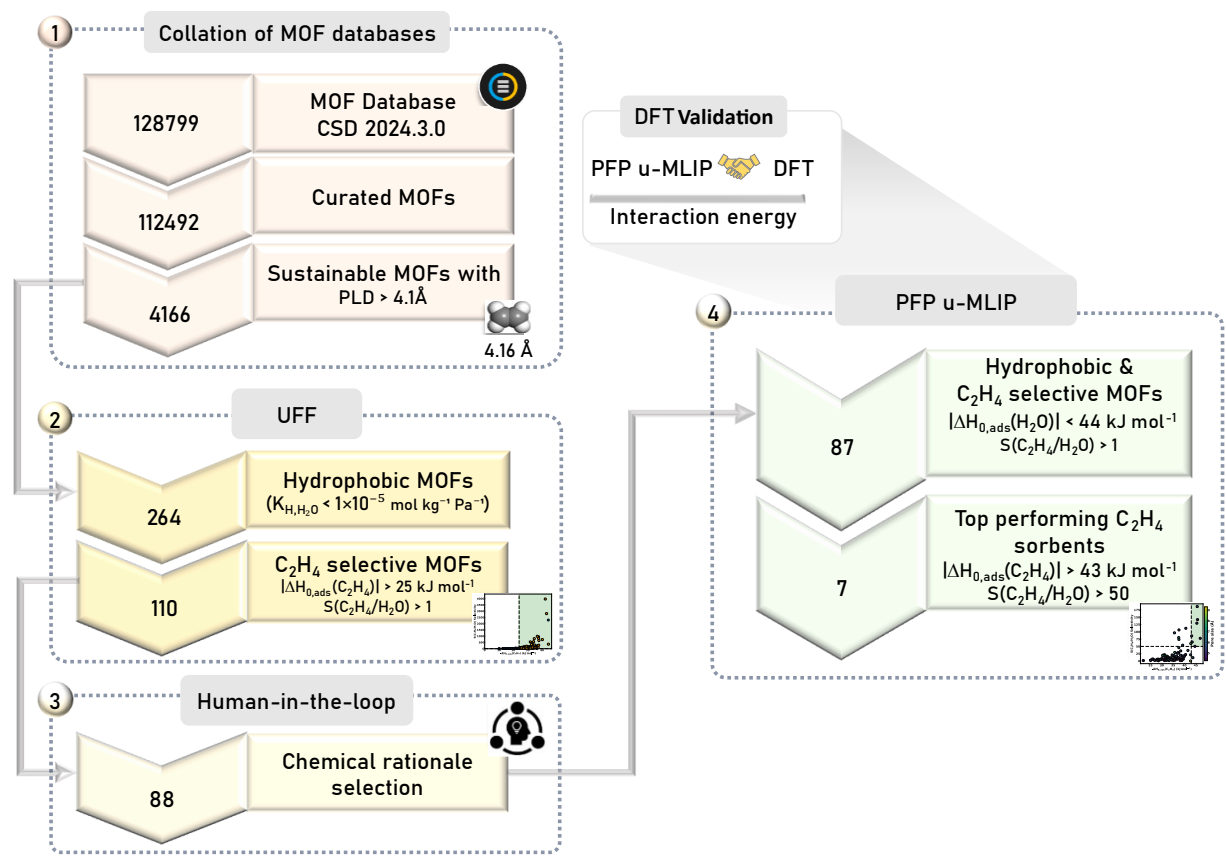

**Figure 1.** Schematic overview of the HTCS workflow for identifying hydrophobic and ethylene-selective MOF candidates from the CSD MOF database (version 2024.3.0). The numbered labels in the top-left corners of each panel indicate the sequential steps in the screening process: **(1)** collation and filtering of MOF structures (PLD > 4.1 Å); **(2)** Evaluation of hydrophobicity and $C_2H_4/H_2O$ ideal selectivity using UFF-based Widom insertion Monte Carlo simulations; **(3)** Human-in-the-loop chemical rationale selection, yielding 88 promising MOFs; **(4)** PFP u-MLIP screening, identifying 87 hydrophobic MOFs [$-\Delta H_{0,ads}(H_2O)$ < 44 kJ mol$^{-1}$] and confirming 7 top-performing sorbents with both high affinity [$-\Delta H_{0,ads}(C_2H_4)$ > 43 kJ mol$^{-1}$] and high selectivity $S(C_2H_4/H_2O)$ > 50]. u-MLIP predictions were benchmarked against DFT adsorption energies as a supporting validation step to confirm model fidelity.

### II.1 Collation and curation of a sustainable MOFs structure database

A high-quality and structurally consistent MOFs database is a prerequisite for reliable HTCS of MOF properties.[57] As such, a systematic curation pipeline was implemented starting from the CSD MOF database (version 2024.3.0)[55] as outlined in Figure 1. We cross-referenced the initial CSD entries with two well-established molecular simulations-ready MOFs: QMOF[58] and MOSAEC-DB.[59] To avoid redundant, time- and computation-intensive curation steps, we excluded CSD MOF entries that overlapped with QMOF (15,966) and/or MOSAEC-DB (71,138), retaining a total of 25,388 MOFs from the non-disordered subset of the CSD MOF database. These MOFs were subsequently curated using SAMOSA (Structural Activation via Metal Oxidation State Analysis) tool,[60] which removes free solvent molecules while preserving chemically relevant monodentate species. This yielded a combined baseline dataset of 112,492 unique MOF structures for our HTCS pipeline. To ensure the accessibility of ethylene (kinetic diameter of 4.163 Å), Zeo++ software was used to filter MOFs with a pore-limiting diameter (PLD) > 4.1 Å. Additionally, MOFs were pre-selected based on their environmentally benign metal ions (Al, Zn, Hf, Fe or Ti) to align with sustainability and health safety standards relevant to food-related applications. Combining these geometric and sustainability criteria yielded a refined subset of 4,166 MOFs.

We then implemented a rigorous refinement process to ensure highest possible chemical correctness and to eliminate structure redundancy. First, we applied the Metal Oxidation State Automated Error Checker (MOSAEC)[61] to identify chemically inconsistent structures by analysing metal oxidation states. MOFs flagged as "BAD" by MOSAEC, typically due to unbalanced charges or chemically implausible valence assignments, were excluded. Next, we addressed structural redundancy by evaluating both topological and geometric similarity using the CoRE-MOF-Tools package.[57] Specifically, we employed the `StructureMatcher` module to identify duplicate or near-duplicate geometries. MOFs with charged frameworks were finally excluded, as these require additional considerations, such as counterion placement, that fall beyond the scope of this study. This multi-stage refinement process yielded a final dataset of 1,881 neutral and computation-ready MOFs that were geometry optimized using the EqV2-ODAC u-MLIP via BFGS optimization in Atomic Simulation Environment (ASE), with a force convergence threshold of 0.05 eV/Å).[62]

**II.2 Monte Carlo Simulations**

*Classical generic FF Widom's test particle insertion method.*

$K_H$ and $\Delta H_{0,ads}$ for both water and ethylene were computed for all EqV2-ODAC u-MLIP optimized MOFs, employing Widom's test particle insertion method at 298 K.[63] Ideal selectivity $S(C_2H_4/H_2O)$ was calculated as the ratio of $K_H$ for ethylene to that for water. Each of these simulations consisted of $2 \times 10^6$ production steps following $1 \times 10^5$ equilibration steps, ensuring statistical convergence. In this calculation, the MOF frameworks were treated as rigid. MOF-guest interactions were modelled as the sum of van der Waals (Lennard-Jones-LJ) and Coulombic contributions. Long-range electrostatic interactions were treated via Ewald summation,[64] and a 12.0 Å cut-off was applied to the van der Waals terms. The UFF parameters[7] were used to define non-bonded terms of the MOF atoms. Atomic partial charges for MOF atoms were assigned using the MEPO-ML (MOF Electrostatic POtential–Machine Learned) model.[65] Ethylene and water molecules were described by a united atom potential model developed by Fisher et al.,[66] and the four center TIP4P-Ew model,[67] respectively. Site–site LJ interactions between unlike force field centers were treated using the Lorentz–Berthelot mixing rules. All these simulations were performed using the Complex Adsorption and Diffusion Simulation Suite (CADSS).[68]

*PFP u-MLIP Widom's insertion simulations.*

The PFP u-MLIP, version 1.19.0[56] was implemented in the Widom's test particle insertion method to assess $K_H$ and $\Delta H_{0,ads}$ for both water and ethylene in the 88 top performing MOFs identified by UFF-based HTCS. To accurately account for dispersion interactions, the PBE-D3 correction[69,70] was applied via `CRYSTAL_U0_PLUS_D3` module, improving the description of the van der Waals interactions between MOFs and the guests. Each simulation was carried out for 50,000 MC cycles. To further investigate the impact of MOF framework flexibility on $\Delta H_{0,ads}$ of ethylene, simulations were carried out using three distinct structural configurations for each of the 88 MOFs: (i) MOF$_{reference}$ - structures optimized (atomic positions relaxed) in the absence of guest molecules using the EqV2-ODAC u-MLIP model; (ii) MOF$_{geometry-opt}$- structures with atomic positions relaxed in the presence of a single ethylene molecule using the PFP u-MLIP; (iii) MOF$_{cell-opt}$-structures with both atomic positions and unit cell parameters relaxed in the presence of a single ethylene molecule using the PFP u-MLIP. The initial structures for (ii) and (iii) were derived from the minimum energy ethylene adsorption configurations identified during the PFP u-MLIP based Widom insertion simulations. These geometry optimizations were performed using the BFGS optimizer [71] with a force convergence threshold of 0.005 eV.Å$^{-1}$. All Widom insertion and geometry optimization calculations using the PFP u-MLIP were performed via the MATLANTIS platform,[72] under an academic license. In this framework, the PFP u-MLIP is implemented through the `Estimator` module interfaced with the `ASECalculator` enabling seamless integration with the Atomic Simulation Environment (ASE) for energy and force evaluations.

**II.3 PFP u-MLIP Benchmarking against DFT calculations**

To validate the accuracy of the PFP u-MLIP used in this study for modelling MOF–guest interactions, specifically in the case of ethylene and water, the current release of the PFP u-MLIP model (version 6.0.0[56]) was rigorously benchmarked against 0 K DFT calculations. The analysis focused on ethylene and water adsorption in 88 top-performing MOFs, with guest positions taken from the lowest-energy adsorption configurations identified via PFP-based Widom insertion MC simulations on MOF$_{reference}$. Each resulting MOF+guest configuration, containing a single guest molecule per simulation box, was then subjected to PFP u-MLIP and DFT geometry optimization, with only the atomic positions relaxed.

The MOF–guest interaction energies ($\Delta E$) for ethylene and water were then evaluated using the following expression:

$$\Delta E = E(\text{MOF} + \text{guest}) - E(\text{MOF}) - E(\text{guest})$$

where, $E(\text{MOF} + \text{guest})$, is the total energy of the geometry optimized MOF–guest configurations, while $E(\text{MOF})$ and $E(\text{guest})$ are the single-point energies of the MOF framework for which the guest molecule is removed and of the isolated guest molecule (ethylene or water), respectively. This approach was applied consistently across both PFP u-MLIP and DFT calculations to ensure a fair comparison.

DFT calculations were carried out using the Quickstep module in the CP2K package.[73] We used the Perdew–Burke–Ernzerhof (PBE)[74] exchange-correlation functional within the Gaussian and plane-wave (GPW) formalism. Double-zeta valence polarized MOLOPT basis sets (DZVP-MOLOPT)[75] and Goedecker–Teter–Hutter (GTH) pseudopotentials[76] were used. Long-range dispersion interactions were included via Grimme's D3 correction[69], consistent with the treatment employed in the PFP u-MLIP framework. A density cutoff of 450 Ry was applied and the geometry optimizations were performed until maximum residual forces were less than $2 \times 10^{-4}$ au·Å$^{-1}$.

## III. Results and discussion

**UFF-guided identification of MOFs for ethylene capture under humidity**

Our multi-stage screening pipeline started with 1,881 charge-neutral, computation-ready MOFs, validated via MOSAEC and CoRE MOF protocols and geometry-optimized at the u-MLIP (EqV2-ODAC) level. The hydrophobic character of these MOFs was first assessed by computing their $K_{H,H_2O}$ using Widom's test particle insertion MC simulations with UFF. Using ZIF-8 (a well-known hydrophobic MOF with $K_{H,H_2O} \approx 2.5 \times 10^{-6}$ mol·kg$^{-1}$·Pa$^{-1}$,[77]) as a reference, we applied a conservative threshold for $K_H = 1 \times 10^{-5}$ mol·kg$^{-1}$·Pa$^{-1}$, yielding a subset of 264 hydrophobic MOFs suitable for ethylene capture in moisture-rich environments.

The ethylene/water selective adsorption of this subset of hydrophobic MOFs was then assessed by calculating their ideal selectivity values $S(C_2H_4/H_2O)$ (=$K_{H,C_2H_4}/K_{H,H_2O}$). MOFs with $S(C_2H_4/H_2O)>1$ were considered as promising sorbents for ethylene adsorption over water. Additionally, to ensure practical relevance, we imposed a thermodynamic constraint, requiring the adsorption enthalpy of ethylene ($\Delta H_{0,ads}(C_2H_4)$) to exceed 25 kJ mol$^{-1}$ (in absolute value) significantly higher than ethylene's vaporization enthalpy (14 kJ mol$^{-1}$).[78] This criterion ensured strong host-guest interactions, critical for trace ethylene capture. These combined filtering, hydrophobicity, $S(C_2H_4/H_2O)>1$ and high $C_2H_4$ affinity, narrowed the list of potential candidates to 110 high-performance MOFs. These materials, which collectively satisfy all three criteria, are highlighted in the shaded green region of **Figure 2a**. To ensure practical applicability, we rigorously evaluated the synthetic feasibility of linkers in the 110 candidate MOFs using the MOFid toolkit[79]. Through systematic filtering and manual inspection (**Figure S1**), we excluded MOFs containing excessively long linkers (backbone lengths >3 nm), highly complex architectures including polycyclic or sterically hindered structures, and non-commercially available linkers. This comprehensive screening process eliminated 22 impractical candidates, yielding a final set of 88 synthetically viable MOFs highlighted in orange colour in **Figure 2a**. **Figure 2b** shows that these selected MOFs predominantly feature moderate pore sizes ranging from 4.5 to 6.0 Å underscoring the importance of pore confinement in promoting selective ethylene adsorption over water.

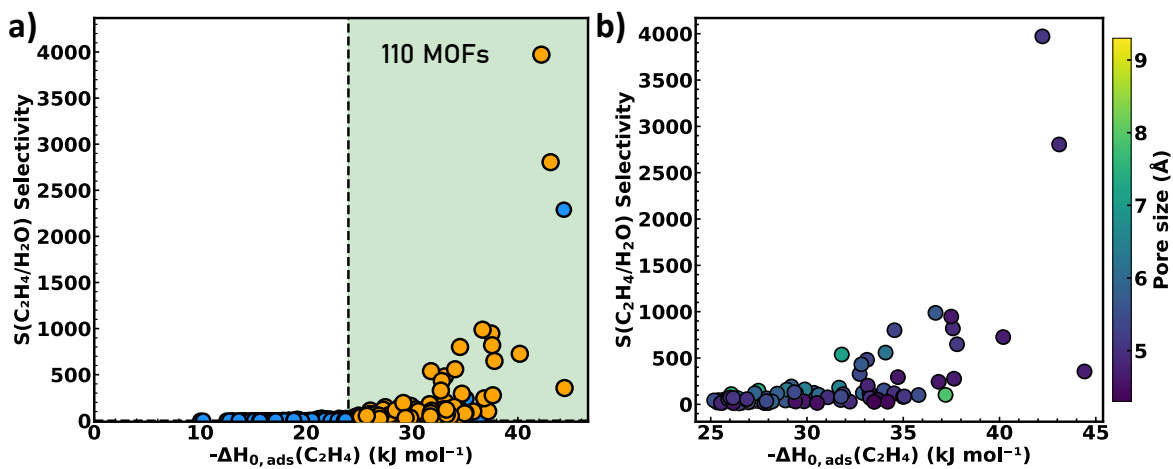

**Figure 2. Initial classical generic FF-based HTCS.** (a) S(C$_2$H$_4$/H$_2$O)-negative adsorption enthalpy of ethylene (-ΔH$_{0,ads}$(C$_2$H$_4$)) plot across the dataset of 264 hydrophobic MOFs. The shaded green region highlights 110 MOFs that exhibit both high ethylene affinity (- ΔH$_{0,ads}$(C$_2$H$_4$) > 25 kJ mol$^{-1}$) and S(C$_2$H$_4$/H$_2$O) > 1, the 88 identified viable MOFs being represented in orange colour; (b) Same plot for this selected sub-set of 88 MOFs alongside their pore sizes.

*Benchmarking u-MLIP against DFT calculations*

We first validated the applicability of PFP u-MLIP to our test case by benchmarking its predicted interaction energies (ΔE) for both C$_2$H$_4$ and H$_2$O against the values obtained by DFT calculations across the representative set of 88 MOFs. **Figure 3a and 3b** emphasize an excellent agreement for both ethylene and water with very low MAD of 2.4 kJ mol$^{-1}$ and 3.0 kJ mol$^{-1}$ respectively. Notably, 91% of the investigated MOFs exhibit deviations within ±5 kJ mol$^{-1}$ for ethylene, while 78% MOFs fall within the same range for water. The remaining MOFs show moderate deviations of 5-10 kJ mol$^{-1}$. **Figure 3c** presents a comparison of the computational costs associated with DFT and PFP u-MLIP calculations across the full benchmarking dataset of 88 MOFs, which have an average system size of 366 atoms (**Figure S2**). This benchmarking also confirms PFP u-MLIP's quantitative reliability in modelling MOF–guest interactions for both polar (H$_2$O) and non-polar (C$_2$H$_4$) guests across diverse MOF frameworks.

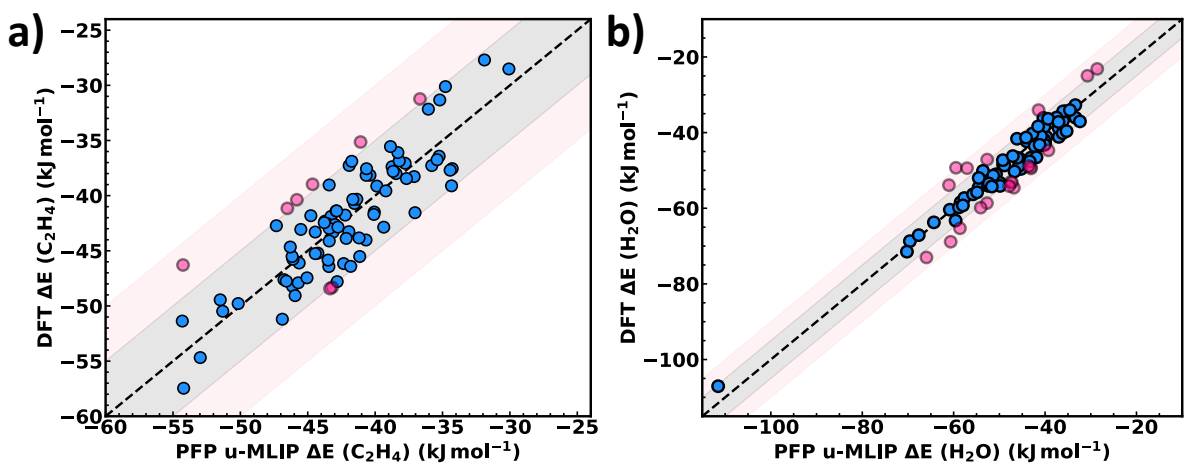

**Figure 3. PFP u-MLIP benchmarking against DFT calculations**. Comparison between interaction energies of (a) ethylene and (b) water for a dataset of 88 MOFs. The dashed black line indicates the ideal parity line. Blue data points within the shaded gray region denote a mean absolute deviation of ±5 kJ mol$^{-1}$, while pink data points within the light pink region indicate deviations between ±5 and ±10 kJ mol$^{-1}$.

*Comparison of classical generic FF and PFP u-MLIP adsorption predictions*

To evaluate the reliability of the conclusions drawn from our initial UFF-based HTCS, we conducted a systematic comparison of adsorption energetics predicted by UFF and those calculated by PFP u-MLIP across the final set of 88 MOF structures. As a critical first validation step, we examined whether PFP u-MLIP confirms the hydrophobic character of these MOFs by computing their $\Delta H_{0,ads}(H_2O)$ under the same rigid MOF framework approximation applied in the initial UFF calculations. As shown in **Figure 4a**, u-MLIP predicts $\Delta H_{0,ads}(H_2O)$ values below the water vaporization enthalpy threshold ($\Delta H_{vap}(H_2O) \approx 44$ kJ mol$^{-1}$) for 87 out of 88 MOFs, indicating weak water-MOF affinity and thereby supporting the hydrophobic classification originally established using UFF, which predicted all values below 30 kJ mol$^{-1}$. Although UFF does not quantitatively agree with the adsorption enthalpy values computed with PFP u-MLIP, it effectively captures the qualitative trend, demonstrating its utility in discriminating hydrophobic MOFs in large-scale screening efforts.

The single MOF outlier **ZSTU-3**[80][80] (CSD refcode TOYHOH[85]; marked with a red cross in **Figure 4a**) leads to a PFP u-MLIP calculated -$\Delta H_{0,ads}(H_2O)$ of ~54 kJ mol$^{-1}$, which exceeds $\Delta H_{vap}(H_2O)$, indicating significantly stronger water-MOF framework interactions compared to UFF predictions. The pore architecture of this MOF is shown in **Figure 4b**, where polar −OH moieties are pointing towards a narrow pore space of 5.9 Å × 4.4 Å. Typically, this type of pore confinement plays a dominant role in enhancing water/polar functions interactions, an effect UFF fails to accurately capture, and would require tailored LJ parameters,[81] which is impractical in a HTCS workflow. Analysis of the lowest-energy adsorption configurations deduced from Widom insertion simulations confirms that PFP u-MLIP captures this confinement effect, placing water molecules as close as 2.1 Å to framework atoms near the −OH moieties (**Figure 4c**). In contrast, UFF predicts the nearest interaction at 4.3 Å, occurring at carboxylate oxygen atoms (**Figure 4d**).

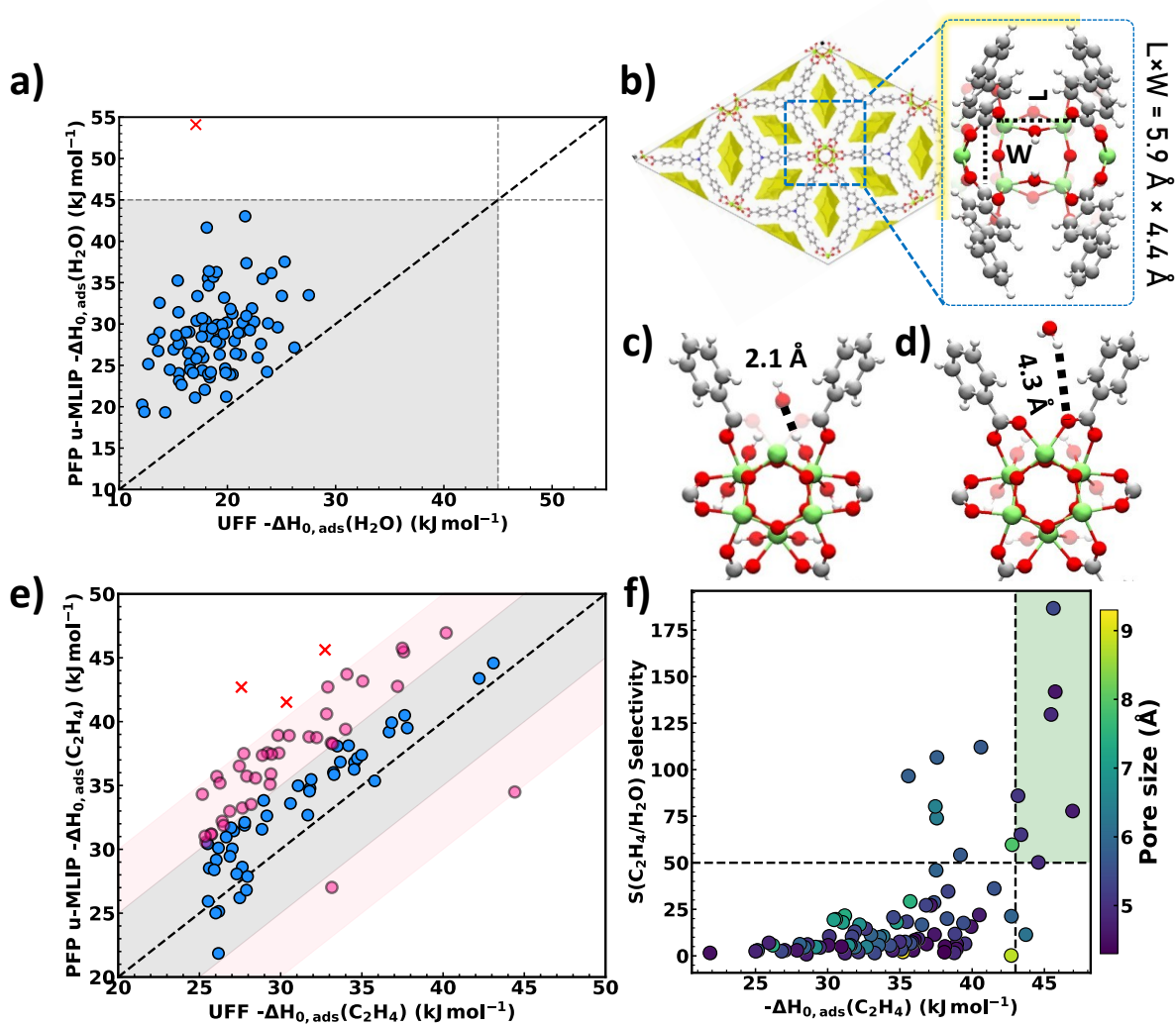

**Figure 4. PFP u-MLIP–based water and ethylene adsorption characteristics across 88 MOF candidates. (a)** Correlation between $\Delta H_{0,ads}(H_2O)$ values calculated using UFF and PFP u-MLIP. MOFs in the grey region exhibit hydrophobic behaviour with $-\Delta H_{0,ads}(H_2O) < 44$ kJ mol$^{-1}$; **(b)** Crystal structure of the outlier MOF, ZSTU-3, with its pore architecture shown as a yellow surface. The inset shows a top view of the confined pore region where −OH groups are enclosed within a narrow pore window of dimensions L (length) = 5.9 Å and W (width) = 4.4 Å; **(c, d)** Water adsorption configurations in ZSTU-3 predicted by **(c)** u-MLIP, showing tight confinement (closest contact distance 2.1 Å), and **(d)** UFF, showing weaker interactions (about 4.3 Å); **(e)** Correlation between $\Delta H_{0,ads}(C_2H_4)$ values from UFF and u-MLIP. Grey and pink shaded regions indicate deviations of <5 and 5-10 kJ mol$^{-1}$, respectively; red crosses mark deviations >10 kJ mol$^{-1}$; **(f)** Correlation between $S(C_2H_4/H_2O)$ and $-\Delta H_{0,ads}(C_2H_4)$ deduced from u-MLIP predictions. Data points are color-coded by pore size.

Following the confirmation of the overall hydrophobic ranking of the MOF candidates, we evaluated their ethylene affinity using both UFF- and PFP u-MLIP-derived $\Delta H_{0,ads}(C_2H_4)$. **Figure 4e** reveals a very good quantitative agreement between both values across all 88 MOFs, with a MAD of 5.1 kJ mol$^{-1}$. This comparative analysis highlights that 52% of the MOFs (blue points) show an excellent correlation between u-MLIP and UFF calculations, with deviations below 5 kJ mol$^{-1}$. Another 44% (pink points), show moderate deviations in the range of 5-10 kJ mol$^{-1}$, while only three MOFs (3%, marked with red crosses), exhibit significant deviations greater than 10 kJ mol$^{-1}$. A closer inspection of these three outlier MOFs: ZTUS-3,[80] A520,[82] and Fe-CFA-6[83] reveals they all contain bridging μ-OH sites confined within angular ('V-shaped') pockets defined by two adjacent linkers (**Figure S3**). In the PFP u-MLIP simulations, ethylene molecules are positioned in such a way to favour short π($C_2H_4$)···OH contacts (~2.5 Å). In contrast, UFF does not capture this interaction mode, instead predicting different ethylene orientations and associated longer host–guest separating distances (3.0-4.1 Å). These differences in predicted geometries illustrate how the choice of potential model can substantially influence the representation of adsorption environments and associated interaction energies in confined pores. Similar to the water interaction modes discussed above, analysis of the ethylene adsorption modes also revealed noticeable differences in the orientation of ethylene molecules between PFP u-MLIP and UFF, underscoring that classical generic force fields may overlook key directional interactions in spatially constrained environments.

We further evaluated the ethylene/water ideal selectivity, S($C_2H_4$/$H_2O$), using $K_{H,C_2H_4}$ and $K_{H,H_2O}$ values obtained from the PFP u-MLIP model (**Figure S4**). The ranking of the top 99% ethylene selective MOFs remains largely consistent with those derived from UFF predictions, with 87 out of the 88 investigated MOFs exhibiting S($C_2H_4$/$H_2O$) > 1. ZSTU-3[80] was the sole exception to exhibit a selectivity value of 0.2 arising from a rather hydrophilic behaviour predicted by PFP u-MLIP simulations as discussed above. In general, the absolute values of S($C_2H_4$/$H_2O$) predicted by the PFP u-MLIP (**Figure 4f**) are consistently lower than those obtained using UFF (**Figure 2b**) mostly due to a more accurate assessment of $K_{H,H_2O}$ with PFP u-MLIP paving the way towards a more realistic estimation of the selectivity.

Further analysis of **Figure 4f** enables to identify among these 87 MOFs, seven top performing candidates combining high S($C_2H_4$/$H_2O$) > 50 and strong ethylene affinity (-$\Delta H_{0,ads}$($C_2H_4$) > 43 kJ mol$^{-1}$). These MOFs are JNU-90 (CSD refcode: NOPGUY[84]), ZnPF-1 (CSD refcode: XEGXUF[85]), EVONOS[86], A520 (CSD refcode: DOYBEA[87]), CYCU-5 (CSD refcode: HONCEU[88]), ATULIM[89], and PERROW[90] (**Figure S5)**. Structurally, these MOFs typically possess microporous cavities in the range of 4.5-5.5 Å, as indicated by the pore size colour gradient represented in **Figure 4f**. The combination of high ethylene selectivity and strong ethylene affinity positions these materials as promising candidates for targeted ethylene capture, particularly in low-trace applications such as food preservation.

**Impact of MOF framework flexibility on ethylene affinity**

Most HTCS workflows inherently assume rigid MOF frameworks, prompting a critical question: to what extent explicit consideration of framework flexibility is essential to refine the prediction of the adsorption properties? To address this question, we equally simulated $\Delta H_{0,ads}$($C_2H_4$) for our shortlisted 88 MOFs via PFP u-MLIP based Widom insertion MC method, applied to geometry optimized MOFs, relaxed either by atomic positions alone or by both atomic positions and unit cell parameters.

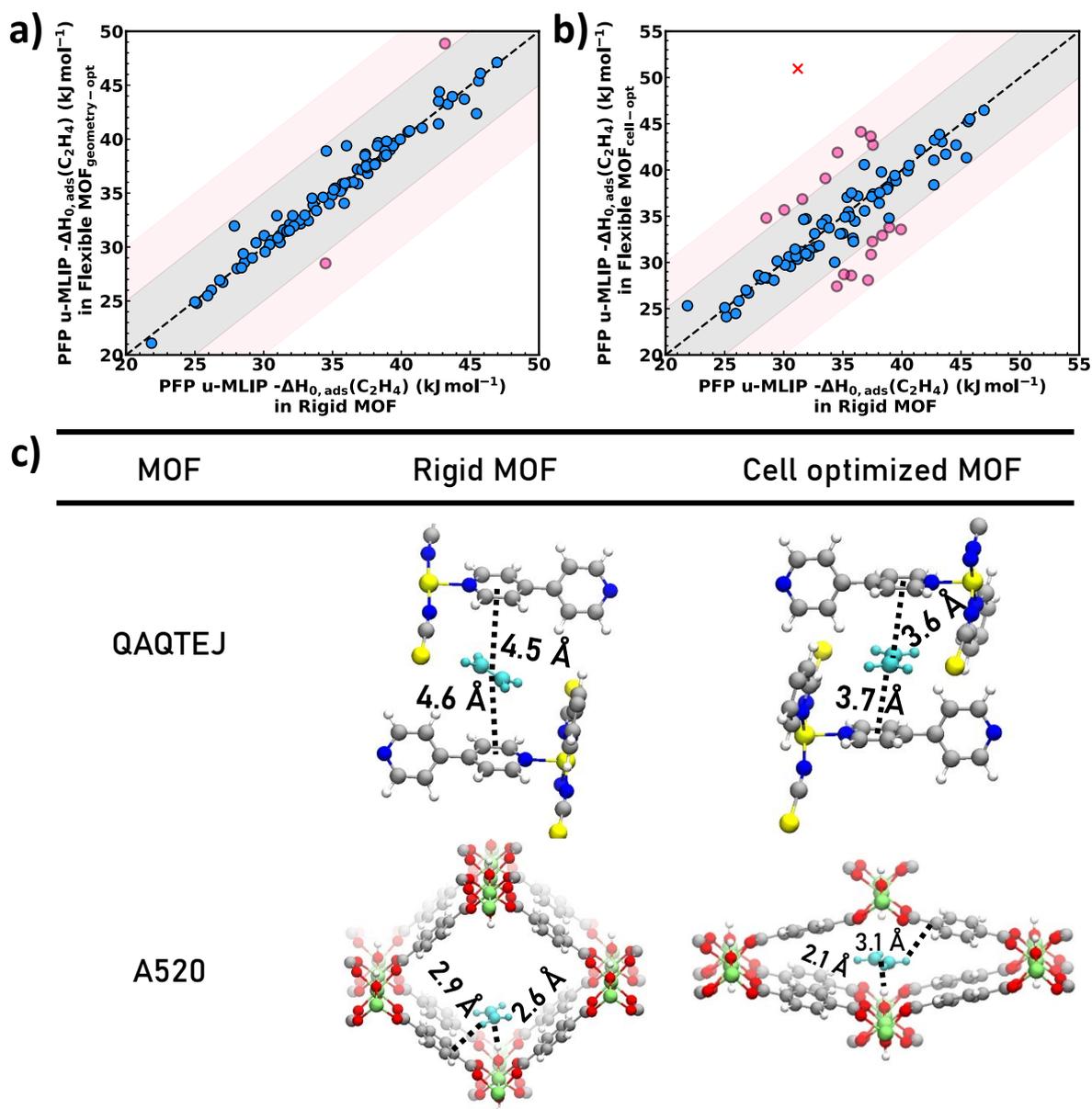

**Figure 5. Impact of MOF framework flexibility on ethylene adsorption energetics**. **(a)** Comparison of $-\Delta H_{0,ads}$ ($C_2H_4$) computed with PFP u-MLIP for rigid versus geometry-optimized (atomic positions only) MOFs. The grey region denotes deviations within ±5 kJ mol$^{-1}$, while the pink region corresponds to deviations between 5 and 10 kJ mol$^{-1}$. **(b)** Comparison of $-\Delta H_{0,ads}$($C_2H_4$) between rigid and fully optimized MOFs (both atomic positions and lattice parameters). The grey and pink regions represent deviations of ≤5 kJ mol$^{-1}$ and 5-10 kJ mol$^{-1}$, respectively. Red cross indicates sole outlier MOF with a deviation exceeding 10 kJ mol$^{-1}$. **(c)** Ethylene adsorption snapshots of QAQTEJ and A520 in both rigid and fully unit cell-optimized configurations. Distances indicate the shortest framework–ethylene contacts.

The comprehensive analyses presented in **Figure 5** reveal that, for majority of the MOFs, the relaxation of their atomic positions exerts only a limited effect on ethylene adsorption energetics, largely preserving the overall ranking as predicted using the rigid MOF framework approximation. Specifically, when only atomic positions are relaxed, 98% of the MOFs (86 out of 88) exhibit $\Delta H_{0,ads}(C_2H_4)$ variations within 5 kJ mol$^{-1}$, with a very low MAD of 0.72 kJ mol$^{-1}$. Only two structures show a more substantial deviation that remains however moderate in the range of 5-10 kJ mol$^{-1}$. This supports those tiny localized atomic rearrangements have negligible influence on ethylene affinity for this MOF set. In the case of full structure optimization (**Figure 5b**), which includes both atomic positions and unit cell parameters relaxation, 80% of MOFs (70 out of 88) still show moderate $\Delta H_{0,ads}(C_2H_4)$ changes within 5 kJ mol$^{-1}$. The remaining 19% exhibit deviations of 5-10 kJ mol$^{-1}$, and just one structure shows a deviation exceeding 10 kJ mol$^{-1}$. Correspondingly, the MAD increases to 2.4 kJ mol$^{-1}$, which however remains modest overall but clearly reflects the larger variability introduced by lattice flexibility for certain MOF frameworks. These results underscore the critical role of lattice flexibility in modulating host–guest interactions, particularly for MOFs whose structural adaptability significantly influences adsorption behaviour. Microscopic insight into these effects is provided in **Figure 5c**, which compares rigid and fully cell-optimized configurations for the outlier MOF QAQTEJ[91] (red cross in **Figure 5b**), a layered structure which displays a distinct flexibility mode: while its internal square nets remain largely intact, unit cell optimization triggers a lateral shift of the layers, reducing the interlayer spacing from 9 to 6 Å. This structural rearrangement results in 35% reduction in unit cell volume and increases -$\Delta H_{0,ads}(C_2H_4)$ from 31 to 51 kJ mol$^{-1}$. The enhancement in adsorption enthalpy can be attributed to increased confinement and stronger $\pi\cdots\pi$ interactions between ethylene and the pyridine rings of the linkers, with contact distances shortening from 4.5 Å to 3.6 Å.

Notably, among the seven identified top-performing MOFs, six of them display small volume changes of 1-4 % leading to a tiny variation of their adsorption enthalpy of ethylene (0.4–4 kJ mol$^{-1}$) (Table S1). In addition, while A520 (CSD refcode: DOYBEA[87]) undergoes a substantial relative volume variation of 19%, yet the change in adsorption enthalpy is only 0.39 kJ mol$^{-1}$. In the relaxed state, the framework retains nearly identical host–guest geometries, with –OH$\cdots\pi$ interactions (2.6 → 2.1 Å) and C–H$\cdots\pi$ contacts (2.9 → 3.1 Å) showing only marginal adjustments, as illustrated in **Figure 5c**.

IV.     Conclusions

This work presents a scalable, multi-scale HTCS framework that couples the speed of classical force fields with the quantum-level accuracy of universal machine learning potentials, validated against DFT, for the identification of high-performance MOF adsorbents for the selective capture of a targeted adsorbate. Using ethylene capture under humid conditions as a challenging benchmark, this approach demonstrates how classical u-IPs and u-MLIPs can be synergistically integrated to achieve both computational efficiency and prediction accuracy for a large diversity of MOFs. Specifically, our results show that classical u-IP methods like UFF-based HTCS remains a robust first-pass filter, correctly capturing adsorption trends in ~97–99% of cases under the rigid-MOF framework approximation. PFP u-MLIPs are indispensable for further refining the adsorption predictions for the top performing MOFs identified from the initial UFF screening especially for MOFs involving strong hydrogen bonding, directional $\pi\cdots H/\pi\cdots\pi$ interactions, or guest-induced framework flexibility, where deviations in $\Delta H_{0,ads}(C_2H_4)$ can reach ~20 kJ mol$^{-1}$. This overall workflow enabled to identify seven top MOF performers combining microporous confinement (4.5-6.0 Å), strong ethylene affinity ($\Delta H_{0,ads}(C_2H_4) > 43$ kJ mol$^{-1}$), and high selectivity ($S(C_2H_4/H_2O) > 50$). These materials directly address the dual challenge of ethylene removal and water tolerance in food storage environment. These computational conclusions are expected to prioritize MOFs for experimental validation and accelerate their integration into scalable postharvest ethylene management technologies. Importantly, our workflow can be applied with any robust u-MLIP, making it broadly applicable. Beyond ethylene capture, the proposed hybrid classical generic FF/u-MLIP strategy is transferable to other adsorption-driven separation, enabling the screening of >100,000 structures while retaining the resolution to treat complex chemistries and flexible frameworks.

**Author contributions**

G.M. conceived the idea and supervised the project. S.B. performed and interpreted the computational calculations. M.W. developed the u-MLIP-based Widom simulation code and contributed to the analysis of the simulated data. S.B. wrote the original draft with input from M.W. and G.M. T.W. and K.H. provided support with the MATLANTIS framework and contributed to manuscript improvement.

**Conflicts of interest**

The authors declare no competing interests.

**Data availability**

The code supporting the PFP u-MLIP–based Widom simulations and the data generated in this study are available on GitHub (https://github.com/gmaurin-group/MLP-WIDOM-SIM).

**Acknowledgements**

The computational work was performed using HPC resources from GENCI-CINES (Grant A0180907613) and MATLANTIS. G.M. thanks Institut Universitaire de France for the Senior Chair.

**Supporting Information**

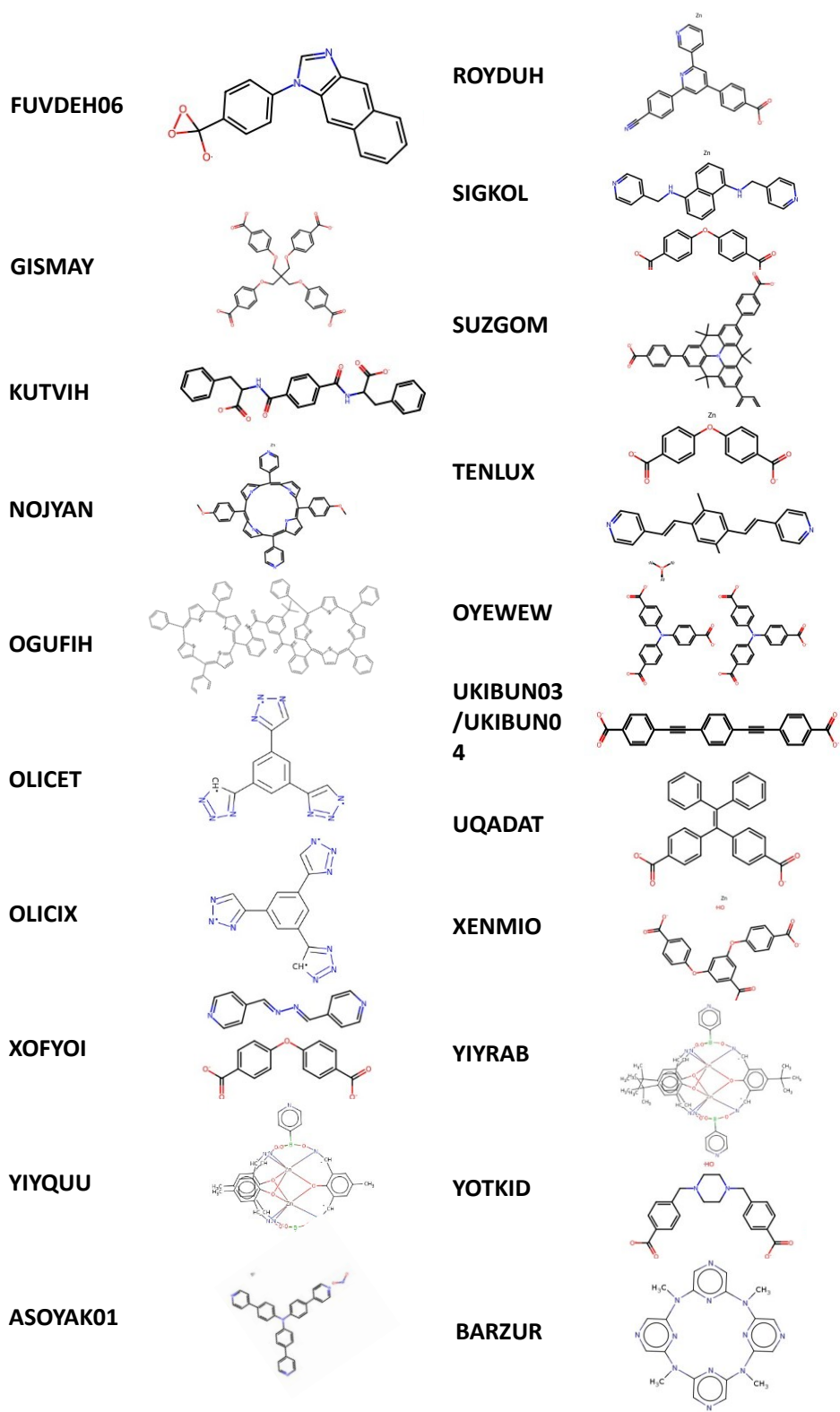

**Figure S1**. CSD refcodes and corresponding linker structures of MOFs evaluated during the synthetic feasibility assessment. The left column lists the MOF CSD refcodes, while the right column illustrates the associated organic linker molecules extracted using the MOFid toolkit. These linkers were analyzed for complexity, and synthetic accessibility.

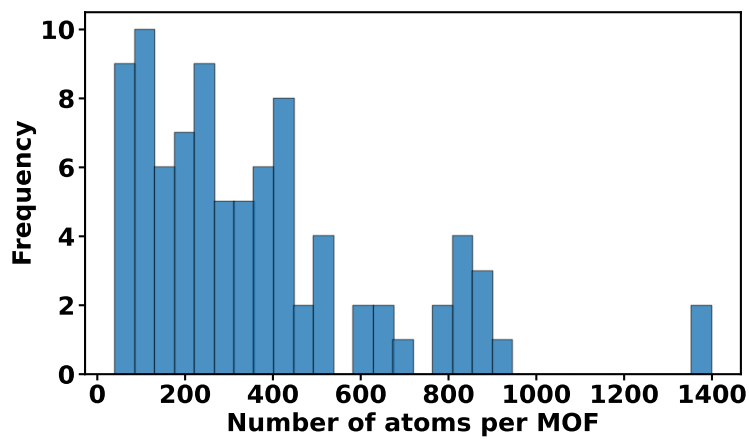

**Figure S2**. Distribution of the number of atoms per MOF in the benchmarking dataset of 88 structures.

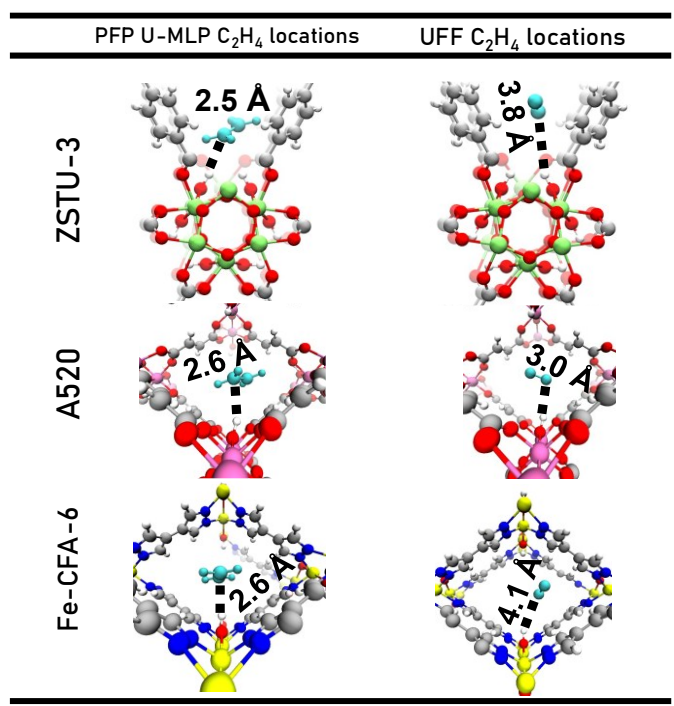

**Figure S3.** Ethylene (cyan) adsorption configurations in three outlier MOFs (red crosses in Figure 4e of the main text), comparing predictions from PFP u-MLIP (left) and UFF (right). Distances indicate the shortest framework–ethylene contacts, highlighting the stronger binding geometries captured by PFP u-MLIP compared to UFF. Hydrogen atoms are omitted in UFF due to the united-atom model.

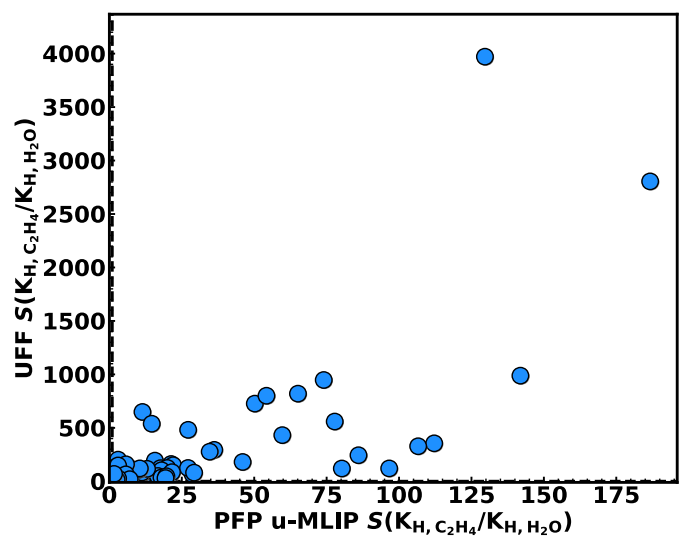

**Figure S4.** Comparison of ethylene/water ideal selectivity $S(C_2H_4/H_2O) = K_{H,C_2H_4}/K_{H,H_2O}$, predicted from UFF and PFP u-MLIP Widom insertion simulations for the identified 88 MOF subset. Each point represents one MOF. Dashed lines indicate $S(C_2H_4/H_2O)=1$.

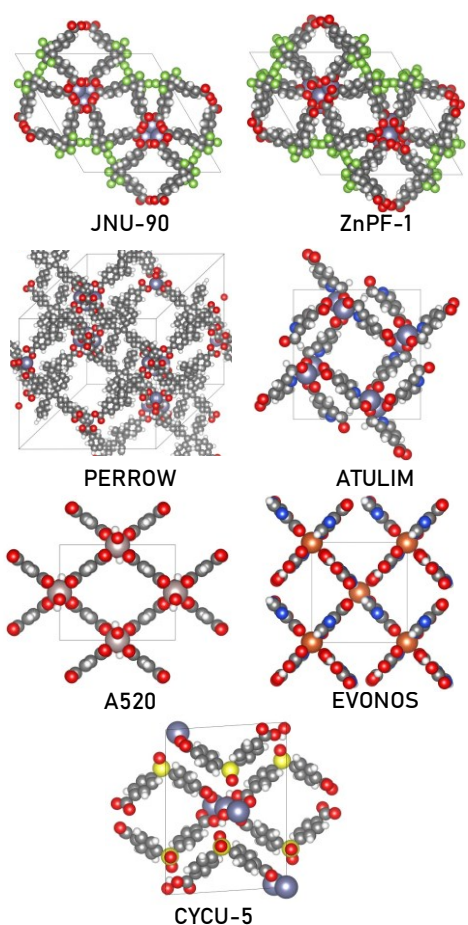

**Figure S5.** Crystal structures of selected top-performing MOFs identified in Figure 4f.

**Table S1.** Adsorption enthalpy and volume variations of the seven top-performing MOFs identified in Figure 4f. Shown are the differences in ethylene adsorption enthalpies ($\Delta(\Delta H_{0,ads}(C_2H_4))$) between rigid and cell-optimized MOF structures, together with the corresponding percentage volume change ($\Delta V\%$). $\Delta(\Delta H_{0,ads}(C_2H_4))$ is defined as $\Delta H_{0,ads}(C_2H_4)$ of the rigid MOF minus $\Delta H_{0,ads}(C_2H_4)$ of the cell-optimized MOF. $\Delta V\%$ is calculated as: $\Delta V\% = V(\text{rigid MOF}) - V(\text{cell-optimized MOF}) \times 100 / V(\text{rigid MOF})$

| MOF | $\Delta(\Delta H_{0,ads}(C_2H_4))$ (kJ·mol$^{-1}$) | $\Delta V\%$ |
|---|---|---|
| JNU-90 | -0.32 | -2.2 |
| ZnPF-1 | -1.88 | -4.3 |
| PERROW | 0.68 | 0.2 |
| ATULIM | -0.23 | -1.6 |
| A520 | -0.39 | 19 |
| EVONOS | -4.13 | 2.2 |
| CYCU-5 | -0.49 | 1.5 |

## Notes and references